\documentclass[pre,twocolumn,showpacs]{revtex4-1}

\newcommand{\Eq}[1]{Eq.~(\ref{eq:#1})}
\newcommand{\Eqs}[1]{Eqs.~(\ref{eq:#1})}
\newcommand{\f}{\mathbf{f}}
\newcommand{\Fig}[1]{Fig.~\ref{fig:#1}}
\newcommand{\Figure}[1]{Figure~\ref{fig:#1}}
\newcommand{\Figs}[1]{Figs.~\ref{fig:#1}}
\newcommand{\Ref}[1]{Ref.~\cite{#1}}
\renewcommand{\O}{{\cal O}}
\newcommand{\Sec}[1]{Sec.~\ref{sec:#1}}
\newcommand{\rr}{\mathbf{r}}
\renewcommand{\v}{\mathbf{v}}

\newcommand{\expt}[1]{\left< #1\right>}
\newcommand{\gdot}{\dot{\gamma}}
\newcommand{\taudiss}{\tau_\mathrm{diss}}
\renewcommand{\u}{{\bf u}}
\usepackage{graphicx}
\usepackage{bm}

\begin{document}
\title{Relaxation Times and Rheology in Dense Athermal Suspensions}

\author{Peter Olsson}
%\email{Peter.Olsson@tp.umu.se}

\affiliation{Department of Physics, Ume\aa\ University, 
  901 87 Ume\aa, Sweden}

\date{\today}   

\begin{abstract}
  We study the jamming transition in a model of elastic particles under shear at zero
  temperature. The key quantity is the relaxation time $\tau$ which is obtained by
  stopping the shearing and letting energy and pressure decay to zero. At many different
  densities and initial shear rates we do several such relaxations to determine the
  average $\tau$. We establish that $\tau$ diverges with the same exponent as the
  viscosity and determine another exponent from the relation between $\tau$ and the
  coordination number. Though most of the simulations are done for the model with
  dissipation due to the motion of particles relative to an affinely shearing substrate
  (the RD$_0$ model), we also examine the CD$_0$ model, where the dissipation is instead
  due to velocity differences of disks in contact, and confirm that the above-mentioned
  exponent is the same for these two models. We also consider finite size effects on both
  $\tau$ and the coordination number.
\end{abstract}

\pacs{63.50.Lm,	%	Equations of state, phase equilibria, and phase
                %	transitions: Glasses and amorphous solids,  
  45.70.-n	%	Classical mechanics of discrete systems: Granular systems
  83.10.Rs 	%	Rheology: Computer simulation of molecular and particle
                %	dynamics
}
\maketitle

\section{Introduction}

Granular materials, supercooled liquids, and foams are examples of systems that may
undergo a transition from a liquid-like to an amorphous solid state as some control
parameter is varied. It has been hypothesised that the transitions in these strikingly
different systems are controlled by the same mechanism \cite{Liu_Nagel} and the term
jamming has been coined for this transition.  Much work on jamming has focused on a
particularly simple model, consisting of frictionless spherical particles with repulsive
contact interactions at zero temperature \cite{OHern_Silbert_Liu_Nagel:2003}. The packing
fraction (density) $\phi$ of particles is then the key control parameter.  Many
investigations have focused on jamming upon compression, and jamming by relaxation from
initially random states \cite{OHern_Silbert_Liu_Nagel:2003,
  Chaudhuri_Berthier_Sastry,Vagberg_VMOT:jam-fss}.  Another, physically realizable and
important case, is jamming upon shear deformation.  This has been modeled with elastic
particles both with a finite constant shear strain rate $\gdot$
\cite{Olsson_Teitel:jamming,Hatano:2008, Hatano:2008:arXiv, Otsuki_Hayakawa:2009b,
  Hatano:2009, Hatano:2010, Tighe_WRvSvH}, and by quasistatic shearing
\cite{Vagberg_VMOT:jam-fss, Heussinger_Barrat:2009,
  Heussinger_Chaudhuri_Barrat-Softmatter}, in which the system is allowed to relax to its
local energy minimum after each finite small strain increment.  A nice method to do
shearing simulations of hard disks has also recently been
developed\cite{Lerner-PNAS:2012}.

Several open questions remain in spite of much studies of the jamming models under steady
shear. Central among them is an understanding of the mechanisms behind jamming, a question
that has been addressed, for the case of hard disks, in several papers by Wyart and
co-workers\cite{Lerner-PNAS:2012, During-Lerner-Wyart:2014, DeGiuli:2014}. A related
question is what details of the models that are important for the universality class. It
has earlier been claimed\cite{Tighe_WRvSvH} that a more realistic model for the
dissipation---where the dissipation is due to the velocity differences between disks in
contact, the CD$_0$ model---gives a different critical behavior than the simpler RD$_0$
model in which the dissipation is against an affinely shearing substrate. Evidence agains
this claim has recently been given in\cite{Vagberg_OT:jam-cdrd}, but much work remains to
clarify other aspects of the various models that are relevant for different physical
systems close to jamming.

In this work we perform large scale simulations to determine the relaxation time---a
quantity whose divergence, we will argue, lies behind the jamming transition. We do that
by first shearing at a steady shear rate and then stopping the shearing and letting energy
and pressure decay to zero; the relaxation time is the time constant of this exponential
decay. We also determine a related time---the dissipation time---which is the time scale
of the initial decay just after stopping the shearing. We characterize the dependencies of
these relaxation times on both distance from (below) jamming and the initial shear
rate. We then motivate a direct relation between the relaxation time and the lowest
vibrational frequency of Lerner et al.\cite{Lerner-PNAS:2012}. Following Lerner et
al.\cite{Lerner-PNAS:2012} we determine the contact number $z$ in the absence of
rattlers. We then find that the relaxation time depends algebraically on the distance to
the isostatic contact number, and determine the exponent for this divergence. Most of our
simulations are for the simpler RD$_0$ model (see below) but we also do the same kind of
analysis for the CD model, and confirm\cite{Vagberg_OT:jam-cdrd} that these two models
appear to behave the same. We then turn to two effects that are related to the finite
system sizes: We first show that the ordinary arithmetic averaging can sometimes give
unexpected effects, and then examine how the number of particles in the simulations
affects the spread in contact number and relaxation time.

The organization of this paper is as follows: In Sec.~II we describe our numerical methods
and give a brief summary of some earlier results that are used throughout the paper. In
Sec.~III we first introduce our two key quantities and discuss their differences and
similarities. We then discuss the relation to the vibrational frequencies in a model of
hard disks\cite{Lerner-PNAS:2012}. Also following \Ref{Lerner-PNAS:2012}, we demonstrate a
direct relation to the contact number and show that the determined exponent is the same
for CD$_0$ as for RD$_0$. We also consider the finite size effects. In Sec.~IV we finally
discuss our results, relate them to earlier works, and make some comments. Sec.~V gives a
short summary.

\section{Model and simulations}

\subsection{Simulations}

Following O'Hern \emph{et al.}\ \cite{OHern_Silbert_Liu_Nagel:2003} we use a simple model
of bi-disperse frictionless soft disks in two dimensions with equal numbers of disks with
two different radii in the ratio 1.4. Length is measured in units of the diameter of the
small particles, $d_s$. With $r_{ij}$ the distance between the centers of two particles
and $d_{ij}$ the sum of their radii, the interaction between overlapping particles is
$V(r_{ij}) = (\epsilon/2) \delta_{ij}^2$ with the relative overlap $\delta_{ij} = 1 -
r_{ij}/d_{ij}$.  We use Lees-Edwards boundary conditions \cite{Evans_Morriss} to introduce
a time-dependent shear strain $\gamma = t\gdot$. With periodic boundary conditions on the
coordinates $x_i$ and $y_i$ in an $L\times L$ system, the position of particle $i$ in a
box with strain $\gamma$ is defined as $\rr_i = (x_i+\gamma y_i, y_i)$.  The simulations
are performed at zero temperature.

We consider two different models for energy dissipation.  The CD model (CD for ``contact
dissipation'') is the model introduced by Durian for bubble dynamics in foams
\cite{Durian:1995}, and was also used by Tighe et al.\ \cite{Tighe_WRvSvH}.  Here
dissipation occurs due to velocity differences of disks in contact,
\begin{equation}
  \f^\mathrm{dis}_{\mathrm{CD},i} = -k_d \sum_j (\v_i - \v_j),\qquad\v_i=\dot\mathbf{r}_i.
\end{equation}
In the second model, RD---``reservoir dissipation''---the dissipation is with respect to
the average shear flow of a background reservoir,
\begin{equation}
  \f^\mathrm{dis}_{\mathrm{RD},i} = -k_d (\v_i - \v_\mathrm{R}(\rr_i)),\qquad
  \v_\mathrm{R}(\rr_i)\equiv\gdot y_i\hat x.
\end{equation}
RD was also introduced by Durian \cite{Durian:1995} as a ``mean-field'' \cite{Tewari:1999}
approximation to CD, and is the model used in many early works on criticality in shear
driven jamming \cite{Tewari:1999, Olsson_Teitel:jamming, Andreotti:2012,
  Lerner-PNAS:2012}.  In both cases the equation of motion is
\begin{equation}
  m_i\dot\v_i = \f^\mathrm{el}_i +\f^\mathrm{dis}_i.
\end{equation}
We are here interested in the overdamped limit, $m_i\to0$ \cite{Durian:1995}. In the RD
model it is straightforward to perform simulations with $m=0$. In the CD model we take
$m=1$ which, for the shear rates we are using, turns out to be small enough to be in the
overdamped limit.  We take $\epsilon=1$ and $k_d=1$.  The unit of time is $\tau_0 = d_s
k_d/\epsilon$.

We focus most of our effort, using longer simulation runs at lower shear rates, for the
model RD$_0$, but we also give results for the model CD for comparison. We use $N=65536$
particles, and shear rates down to $\gdot=10^{-9}$ and integrate the equations of motion
with the Heuns method with time step $\Delta t=0.2\tau_0$.

\subsection{Background}

The present paper focuses on the behavior of the above-mentioned models just below
$\phi_J$. We here summarize a few results that are important in the following. 

The jamming transition is a zero-temperature transition from a liquid to a disordered
solid upon the increase of density. An excellent way to probe this transition is to look
at the resistance to shearing. Since the defining property of a liquid is that it is a
material that cannot sustain a shearing force, a finite shear stress, $\sigma$, in the
limit $\gdot\to0$ is a clear sign of a solid phase.  Within the liquid, i.e.\ at
$\phi<\phi_J$, the approach to jamming is seen in the rapid increase of the viscosity,
$\eta=\sigma/\gdot$; numerical evidence suggest that it diverges algebraically,
\begin{equation}
  \label{eq:s-divergence}
  \eta(\phi,\gdot\to0) = \sigma/\gdot \sim (\phi_J-\phi)^{-\beta}.
\end{equation}
Another quantity that clearly signals the transition is the pressure and the pressure
equivalent of the viscosity, $\eta_p=p/\gdot$, which similarly diverges with the exponent
$\beta$,
\begin{equation}
  \label{eq:p-divergence}
  \eta_p(\phi,\gdot\to0) = p/\gdot \sim (\phi_J-\phi)^{-\beta}.
\end{equation}
Since $p\sim\delta$ whereas the interaction energy is $E\sim\delta^2$, the energy
diverges with the exponent $2\beta$,
\begin{equation}
  \label{eq:E-divergence}
  \lim_{\gdot\to0} E/\gdot^2 \sim (\phi_J-\phi)^{-2\beta}.
\end{equation}
Equations (\ref{eq:s-divergence}) and (\ref{eq:p-divergence}) for $\sigma$ and $p$, should
hold very close to $\phi_J$, but since the dimensionless friction, $\mu\equiv\sigma/p=\eta/\eta_p$,
has a strong $\phi$-dependence, \Eqs{s-divergence}, (\ref{eq:p-divergence}) clearly give
only the leading divergence of $\eta$ and $\eta_p$, and are not exact expressions that
hold over any finite density interval. To handle this one needs to include \textrm{corrections to
  scaling}\cite{Olsson_Teitel:gdot-scale, Kawasaki_Berthier:2015} by writing
\begin{equation}
  \label{eq:Ogdot-scale-beta}
  \O/\gdot \sim (\phi_J-\phi)^{-\beta} [1 + c_\O (\phi_J-\phi)^{\omega\nu}],
\end{equation}
for the observables $\sigma$ and $p$.

In simulations of soft particles the data will depend on the shear rate, $\gdot$, which
may be considered a relevant scaling variable. This suggests a scaling assumption as in
critical phenomena\cite{Olsson_Teitel:jamming}. With $\delta\phi=\phi-\phi_J$,
\begin{equation}
  \O(\delta\phi, \gdot) = b^{-y/\nu} g_\O(\delta\phi b^{1/\nu}, \gdot b^z),
  \label{eq:O-scale}
\end{equation}
where $b$ is typically considered to be a length rescaling factor, though it can be chosen
arbitrarily.  With $b=|\delta\phi|^{-\nu}$, specializing to $\delta\phi<0$, the scaling
relation for $\O/\gdot$ becomes
\begin{equation}
  \label{eq:Ogdot-scale}
  \O(\delta\phi, \gdot)/\gdot = |\delta\phi|^{-(z\nu-y)} g_\O(\gdot/|\delta\phi|^{z\nu}).
\end{equation}
In the $\gdot\to0$ limit $g_\O(x\to0)=$ const, together with \Eq{Ogdot-scale-beta} leads
to the identification $\beta=z\nu-y$. 

Corrections to scaling are included by generalizing \Eq{O-scale} to
\begin{equation}
  \label{eq:O-scale-corr}
  \O = b^{-y/\nu} [g_\O(\delta\phi b^{1/\nu}, \gdot b^z) + b^{-\omega} h_\O(\delta\phi b^{1/\nu}, \gdot b^z)],
\end{equation}
An analysis based on this kind of approach\cite{Olsson_Teitel:gdot-scale} gave
$\beta=2.77(20)$ whereas a related approach in terms of an effective density
\cite{Olsson_Teitel:jam-HB} gave the very similar $\beta=2.58(10)$. Other recent values
in the literature from simulations are $\beta=2.2$\cite{Andreotti:2012}, and a recent
theoretical work gives $\beta=2.77$\cite{DeGiuli:2014}.

\section{Results}

\subsection{Measured quantities}

\subsubsection{The relaxation time}

\begin{figure}
  \includegraphics[width=7cm]{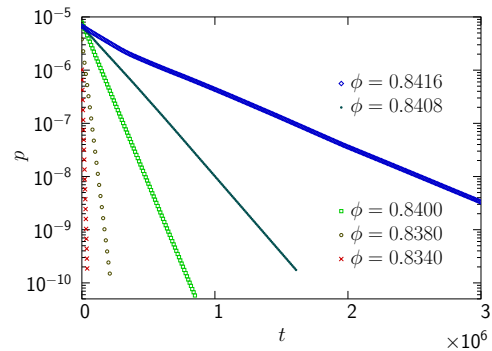}
  \caption{Examples of the pressure relaxation at different $\phi$. The figure shows the
    pressure relaxation after the shearing has been switched off.  The preceeding
    shearings were performed at very low shear rates in order to stay close to the linear
    region; the densities and the initial shear rates were $(\phi,\gdot)=(0.8340,
    10^{-8})$, $(0.8380, 10^{-8})$, $(0.8400, 5\times10^{-9})$, $(0.8408,
    2\times10^{-9})$, $(0.8416, 10^{-9})$.  To determine the relaxation times, $\tau$, we
    fit pressure to an exponential decay, only using data with $p<10^{-7}$.}
  \label{fig:p-time}
\end{figure}

One of the hallmarks of the jamming transition is a diverging time scale. It has been
common to measure this time scale implicitly by measurement of a diverging transport
coefficient like $\eta$ or $\eta_p$.  In this section, however, we measure such a time
scale by looking directly at the relaxation of the system from an initial shear driven
steady state to the zero-energy state obtained after the shearing is turned off.  We
thus make use of a two-stage process: In the first stage the system is driven at steady
shear with a constant shear rate $\gdot$, in the second stage the shearing is stopped but
the dynamics is continued which makes the system relax down to a minimum energy. As the
simulations discussed here are at densities somewhat below $\phi_J$, the final state is
always a state of zero energy, and after a short transient time, energy and pressure decay
exponentially to zero. The relaxation time for a single relaxation is denoted by $\tau_1$,
\begin{displaymath}
  p(t) \sim \exp(-t/\tau_1).
\end{displaymath}
A few such relaxations at different densities are shown in \Fig{p-time}. In each case the
relaxation time is determined from the data with $p(t)<10^{-7}$, where the decay is
exponential to an excellent approximation. As we will see below the relaxation time
depends on the shear rate applied before the relaxation and we will let
$\tau(\phi,\gdot)$---which thus depends on both $\phi$ and $\gdot$---denote the average
relaxation time from about 10--100 such relaxations.

\begin{figure}
  \includegraphics[width=7cm]{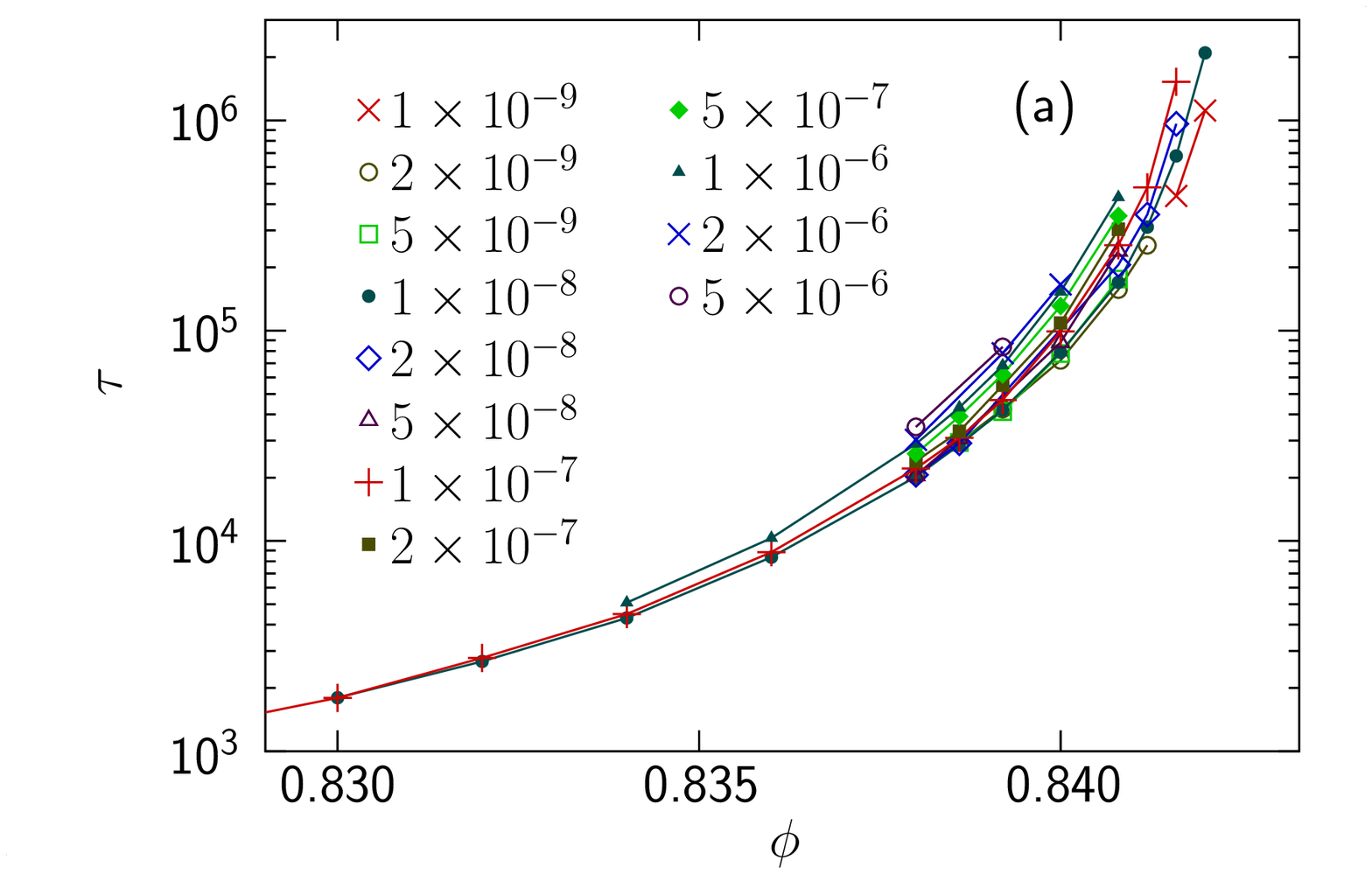}
  \includegraphics[width=7cm]{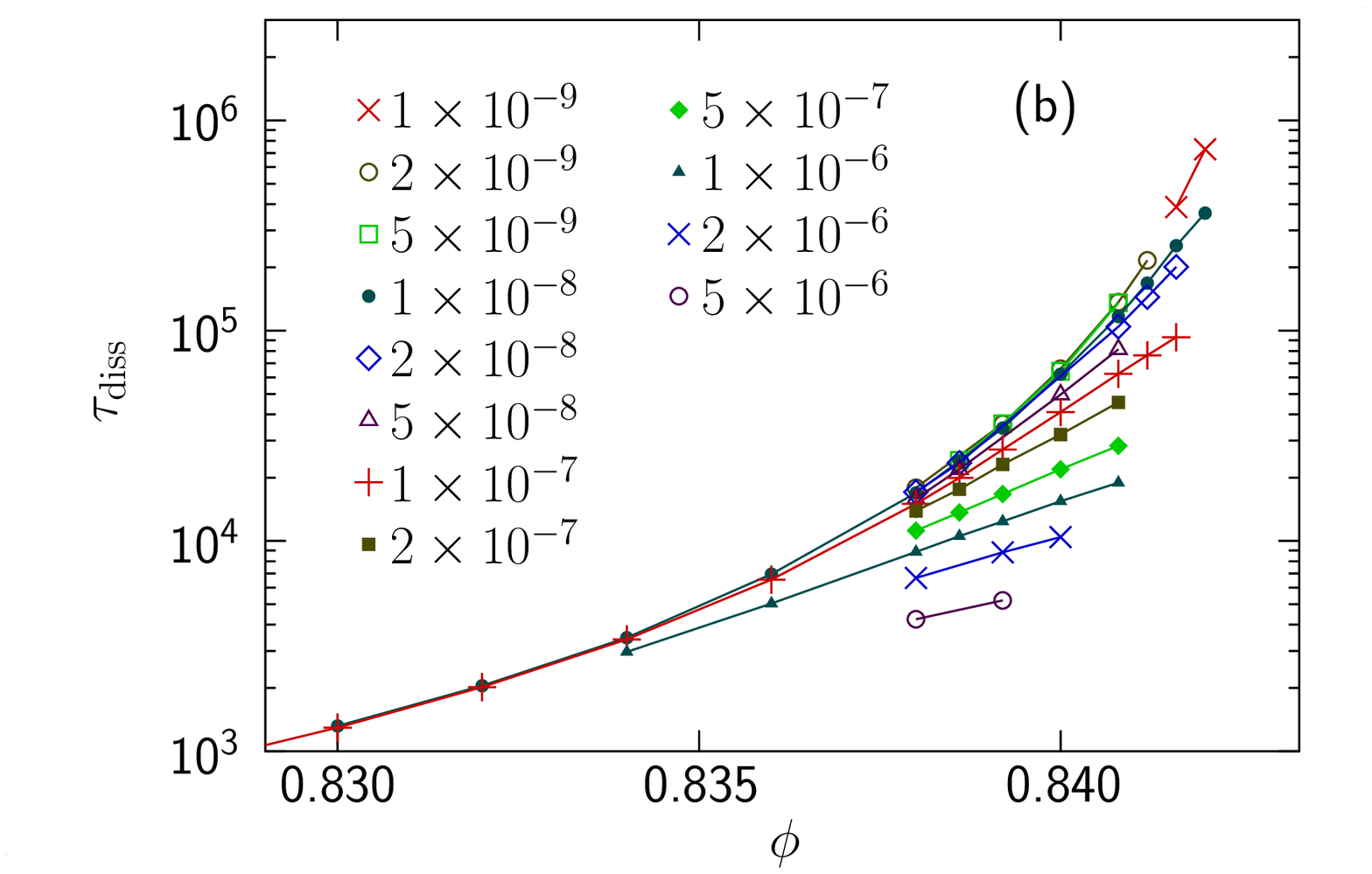}
  \includegraphics[width=7cm]{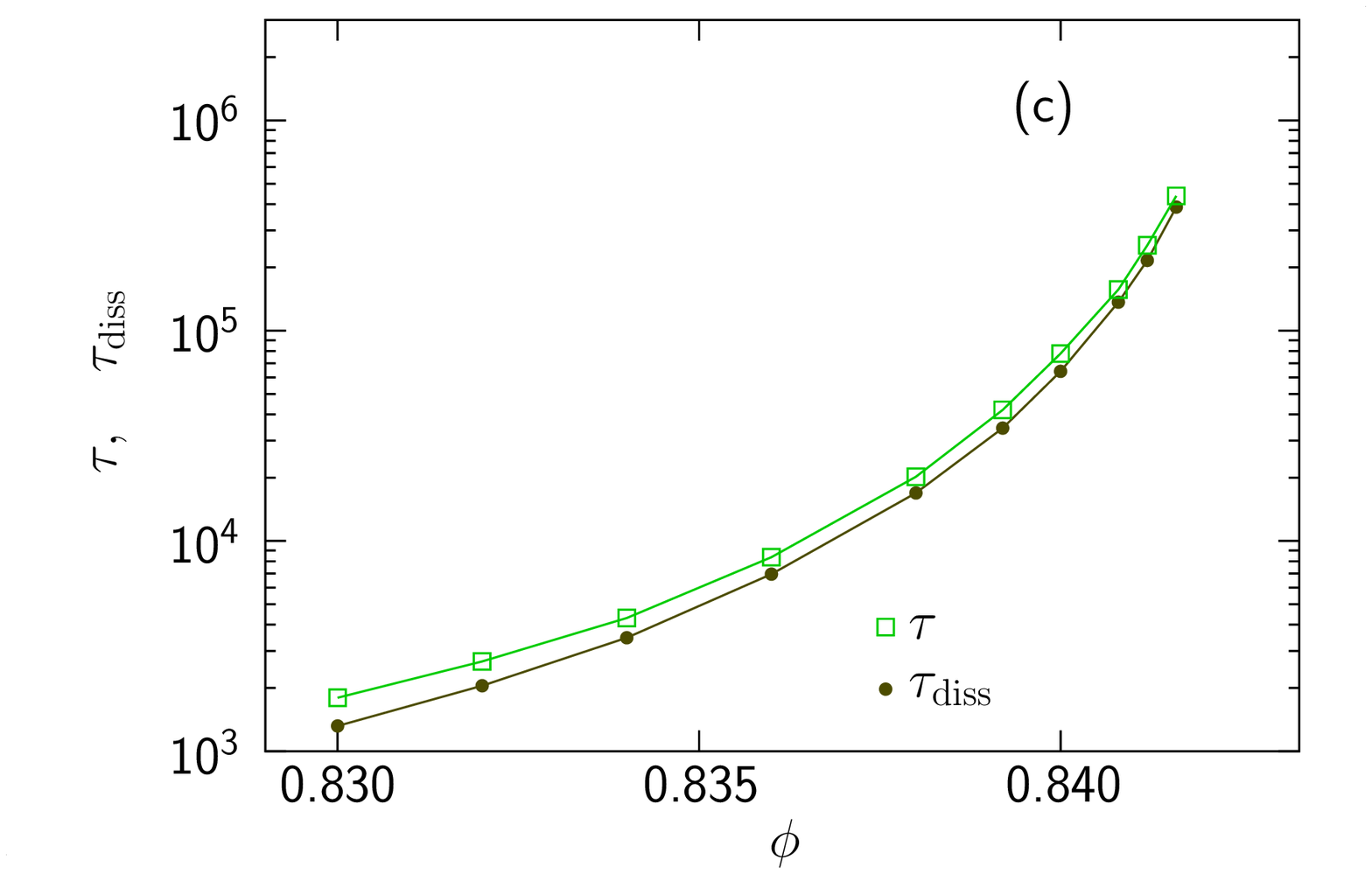}
  \caption{Relaxation time and dissipation time vs density. Panel (a) shows $\tau$ vs
    $\phi$ at several different shear rates. The data increases rapidly with increasing
    $\phi$ suggestive of a divergence at $\phi_J$. There is also a clear shear rate
    dependence, $\tau$ decreases when $\gdot$ is decreased towards the hard disk limit,
    $\gdot\to0$. Panel (b) which shows $\taudiss$ vs $\phi$ also increases rapidly with
    $\phi_J$. The shear rate dependence is however the opposite; $\taudiss$ increases with
    decreasing $\gdot$. Panel (c) shows a comparison of $\tau$ and $\taudiss$ which only
    includes the data with the lowest $\gdot$ (i.e.\ closest to the hard disk
    limit). $\tau$ and $\taudiss$ behave essentially the same across this density
    interval, they are very close at the highest density close to $\phi_J$, but the
    (relative) difference increases with decreasing $\phi$.}
  \label{fig:tau-phi}
\end{figure}

\Fig{tau-phi}(a) which is $\tau(\phi,\gdot)$ versus $\phi$ for several different shear
rates, clearly suggests that $\tau$ diverges at the jamming transition. The figure also
illustrates the shear rate dependence; $\tau$ gets bigger for larger $\gdot$ which means
that the system driven at higher shear rates needs longer time for reaching the
zero-energy state. The reason for this behavior is maybe not entirely obvious, but one can
at least say that the opposite behavior---that the decay were faster for a higher initial
shear rate---would be very counterintuitive.  Recall that this is the shear rate before
the relaxation step; the relaxation itself is performed with $\gdot=0$.

Note also that this relaxation time is a different quantity from the quantity with the
same name in the context of supercooled liquids. In supercooled liquids the particles'
motion is due to the non-zero temperature, whereas the motion in the present context is
due to the relaxation of the potential energy.

% plo p-time.plo

% plo tau-phi.plo

\subsubsection{Dissipation time}

As a complement to the relaxation time, which is determined from the final decay of the
pressure, we also introduce the ``dissipation time'' $\taudiss$, which is defined from the
initial decay rate, just after the shearing has been turned off. For this quantity there
is however no need to study the actual relaxations; at any moment the relaxation rate for
the energy may be determined from the energy together with the dissipating power, giving
$\tau'=E/P_\mathrm{diss}$. In steady shear we may equate the dissipated power with the
input power $P_\mathrm{in}=V\sigma\gdot$, which gives $\tau'=E/(\sigma\gdot)$ for the
average dissipation time. As we want a quantity that may be directly compared to
$\tau$---i.e.\ the decay time for pressure rather than the decay time for energy---we note
that $p\sim\delta$ whereas $E\sim\delta^2$ which means that $p(t) \sim e^{-t/\tau}$
implies $E(t)\sim e^{-t/(\tau/2)}$, and that the two relaxation times differ by a factor
of two. Our final expression for the dissipation time is therefore
\begin{equation}
  \taudiss = 2\frac{E}{\sigma\gdot}.
  \label{eq:taudiss}
\end{equation}

% plo taudiss-phi.plo; plo tau,taudiss-phi.plo
\begin{figure}
  \includegraphics[width=7cm]{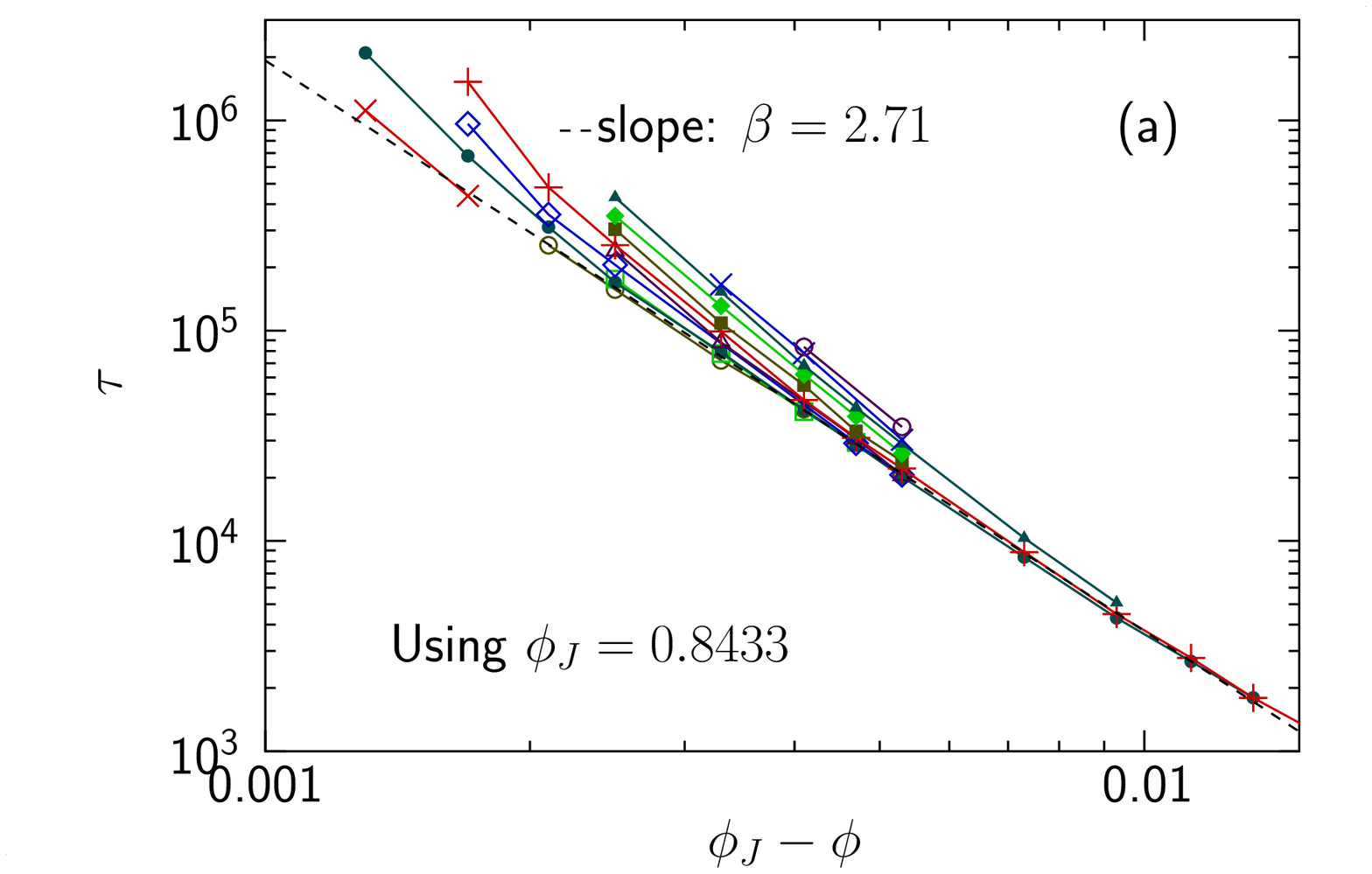}
  \includegraphics[width=7cm]{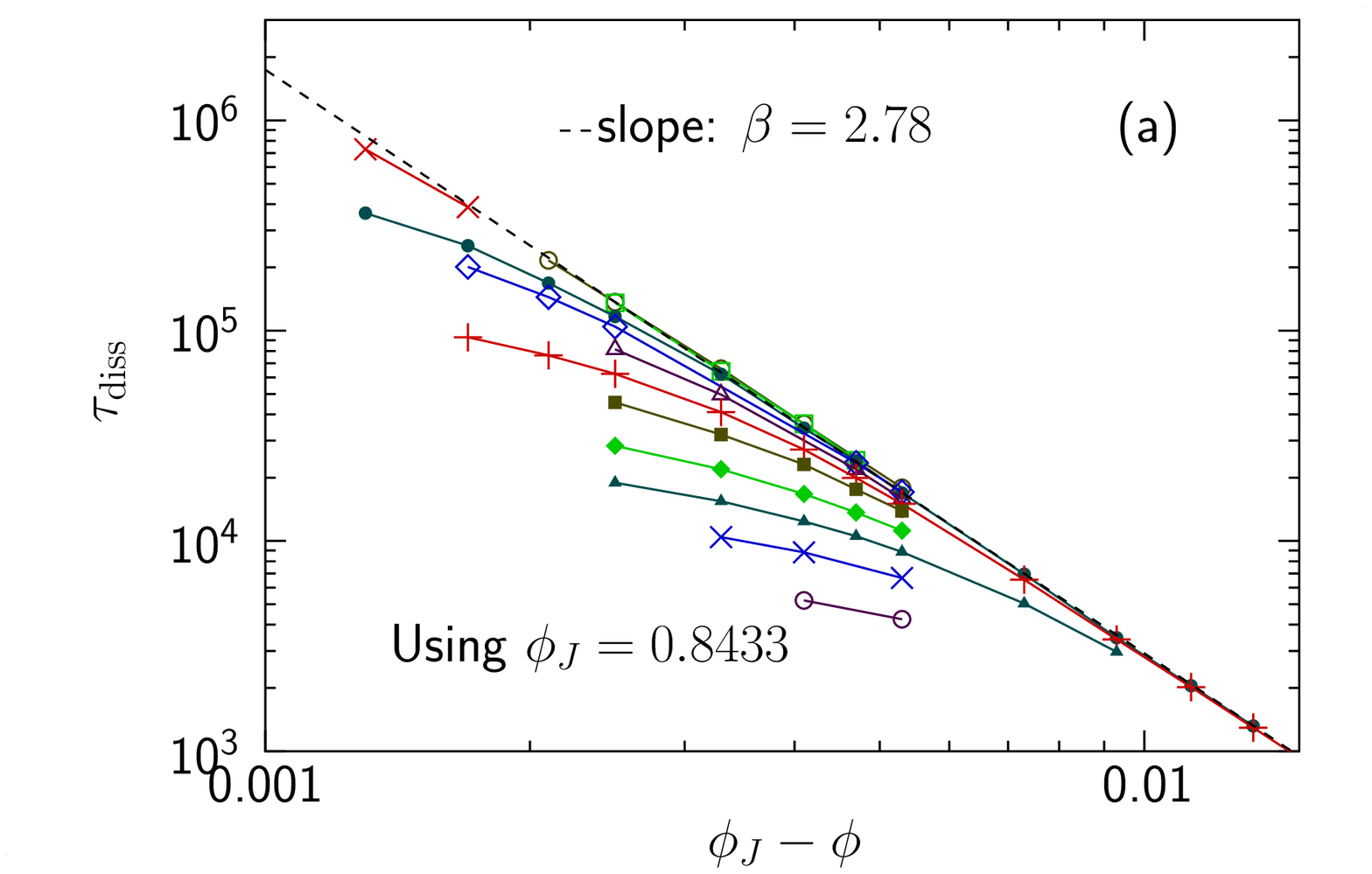}
  \caption{Divergence of $\tau$ and $\taudiss$. We here fix $\phi_J=0.8433$ and determine
    $\beta$ by fitting the few points of $\tau$ and $\taudiss$, respectively, with
    $\phi_J-\phi<0.006$ and sufficiently small $\gdot$ to be close to the hard disk
    limit. (The points for $\phi=0.8420$ and $\gdot=10^{-9}$ appear to be too far from the
    hard disk limit and are not included in the fits.)}
  \label{fig:tau-dphi}
\end{figure}

\Figure{tau-phi}(b) shows $\taudiss$ against $\phi$ for several different shear
rates. Just as for $\tau$ this quantity also appears to diverge as $\phi\to\phi_J$. The
$\gdot$-dependence is however different; $\taudiss$ decreases with increasing $\gdot$,
which means that the relative decrease of the energy is bigger in simulations at higher
shear rates. The different behaviors of $\tau$ and $\taudiss$ is presumably because
$\taudiss$ picks up contributions from all kinds of decay modes, and the faster modes are
more excited when the system is driven with a higher shear rate. In contrast, $\tau$ only
gets contributions from the slowest decay mode.

\Figure{tau-phi}(c) shows a comparison of $\tau$ and $\taudiss$. To eliminate effects due
to the finite shear rate we only include the data at the smallest shear rate and exclude
the data at $\phi=0.8420$, $\gdot=10^{-9}$ which is clearly away from the $\gdot\to0$
limit.

\subsubsection{Divergence}

We are now ready to demonstrate one of the key results of the present paper, which is that
both $\tau$ and $\taudiss$ diverge with the exponent $\beta$.  From the definition of the
dissipation time in \Eq{taudiss} together with \Eqs{E-divergence} and
(\ref{eq:s-divergence}), it follows directly that $\taudiss$ diverges with the exponent
$\beta$:
\begin{equation}
  \label{eq:taudiss-beta}
  \taudiss = \frac{E/\gdot^2}{\sigma/\gdot} \sim
  \frac{(\phi_J-\phi)^{-2\beta}}{(\phi_J-\phi)^{-\beta}} \sim 
  (\phi_J-\phi)^{-\beta}.
\end{equation}
That $\tau$ diverges in the same way follows from the very similar behaviors in
\Fig{tau-phi}(c) but in \Sec{res-hard-p} we will also argue for a direct connection
between $\eta_p$ and $\tau$ by other means.

\Figs{tau-dphi} show the determination of $\beta$ from $\tau$ and $\taudiss$. The
determinations are based on the data points from in \Fig{tau-phi}(c) very close to
$\phi_J$, $\phi_J-\phi<0.006$.  With only a few points with limited precision in a narrow
interval of $\phi$, it is difficult to do a fit with both $\beta$ and $\phi_J$ as free
parameters. We therefore instead determine $\beta$ after fixing the jamming density to
$\phi_J=0.8433$\cite{Heussinger_Barrat:2009, Olsson_Teitel:gdot-scale,
  Olsson_Teitel:jam-HB}. The actual fits of $\tau$ and $\taudiss$ are shown in
\Figs{tau-dphi} and give similar values for the exponent: $\beta=2.71$ and $\beta=2.78$ in
good agreement with earlier estimates\cite{Olsson_Teitel:gdot-scale, Olsson_Teitel:jam-HB,
  DeGiuli:2014}.

\subsection{Relations to hard disk simulations}

In this Section we will relate the relaxation time $\tau$ to results from the study of the
vibrational modes of sheared hard disks\cite{Lerner-PNAS:2012}. Relations between these
two approaches are expected since soft disk simulations at sufficiently low shear rates
give vanishingly small overlaps and therefore should behave just as hard disks.

\subsubsection{Relaxation time and the vibrational frequency}

To motivate the relation between the relaxation time and the vibrational frequency we
consider small displacements $\u_i$ from a zero-energy state. Written in terms of the
vector $\u$, with $2N$ elements, and the stiffness matrix ${\cal M}$, such that the force
(also a vector with $2N$ components) becomes $\epsilon{\cal M} \u$, the equation of motion
for inertial dynamics may be written
\begin{equation}
  \label{eq:vibr}
  m\frac{d^2\u}{dt^2} = \epsilon{\cal M} \u.
\end{equation}
(We here consider a finite mass although our work is concerned with the overdamped limit
of $m \to 0$, only to be able to relate to other approaches.)  With eigenvalues
$\lambda^{(k)}$ and eigenvectors $\u^{(k)}$, the force due to a general displacement
field, $\u=\sum_k c_k \u^{(k)}$ becomes $\epsilon\sum_k \lambda^{(k)} c_k \u^{(k)}$ and
the ansatz $\u(t) = \sum_k c_k \u^{(k)} \sin\omega_k t$ gives $\omega_k^2 = -(\epsilon/m)
\lambda^{(k)}$.  However, below $\phi_J$ where the number of contacts is below the
isostatic value there are modes with zero energy and $\omega_k=0$, which complicates the
analysis. From the formalism for shearing of hard disks Lerner et
al. \cite{Lerner-PNAS:2012} derived a matrix with the same eigenvalues as ${\cal M}$
except for these zero-energy modes. For that matrix the lowest frequency,
$\omega_\mathrm{min}$, is always finite.

The relaxation may similarly be analyzed in terms of small displacements and for
overdamped dynamics the equation of motion becomes
\begin{equation}
  \label{eq:relax}
  k_d\frac{d\u}{dt} = \epsilon{\cal M} \u.
\end{equation}
The ansatz of an exponential decay, $\u(t) = \sum_k \u^{(k)}\exp(-t/\tau_k)$ then gives
$\tau_k^{-1} = -(\epsilon/k_d) \lambda^{(k)}$. Taken together, \Eqs{vibr} and (\ref{eq:relax})
give the desired relation between the relaxation time and the vibrational frequencies,
\begin{equation}
  \tau_k = \frac{k_d}{m} \omega_k^{-2}.
\end{equation}
Our
largest $\tau_k$---the same as our relaxation time, $\tau$---then corresponds to the
lowest frequency, $\omega_\mathrm{min}$.
\begin{equation}
  \label{eq:tau-omega}
  \tau \sim  \omega_\mathrm{min}^{-2}.
\end{equation}
Our observation that there is only a single relaxation time that controls the decay
corresponds well with the finding\cite{Lerner-PNAS:2012} that the lowest frequency in the
vibrational analysis is an isolated mode. If that were not the case, one would expect
several decay modes with similar relaxation times and that would be seen through a
curvature in the data in \Fig{p-time}.

\subsubsection{Relation to pressure}
\label{sec:res-hard-p}

The formalism of \Ref{Lerner-PNAS:2012} gives the relation
\begin{displaymath}
  \omega_\mathrm{min}^{-2} \sim\eta_p,
\end{displaymath}
to be valid in the hard disk limit. Together with \Eq{tau-omega} this leads us
to expect that $\tau$ and $\eta_p$ should behave the same in the hard disk limit and
\Figs{etap,tau} shows comparisons of $\tau$ and $\eta_p$ from our soft disk
simulations with different shear rates. The data clearly approach one another as
$\gdot\to0$.

Panel (a) shows $\tau$ together with $A_p\eta_p$ (where the constant is $A_p=36$) against
$\phi$ for different $\gdot$. Both quantities do indeed appear to approach the same curve
in the $\gdot\to0$ limit, given by the dashed line, $f_\tau(\phi) \sim
(\phi_J-\phi)^{-2.6}$. Panel (b) which shows the same data, but now relative to
$f_\tau(\phi)$, serves as a strong confirmation of the expected equality and gives ample
support for the expected direct proportionality between $\tau$ and $\eta_p$ in the hard
disk limit. Recall that $\eta_p$ and $\tau$ are very different quantities as the first is
determined at constant shearing whereas the second is from the relaxation rate of the
pressure.

% plo etadivtau-s.plo; plo tau,etap-phi.plo
\begin{figure}
  \includegraphics[width=7cm]{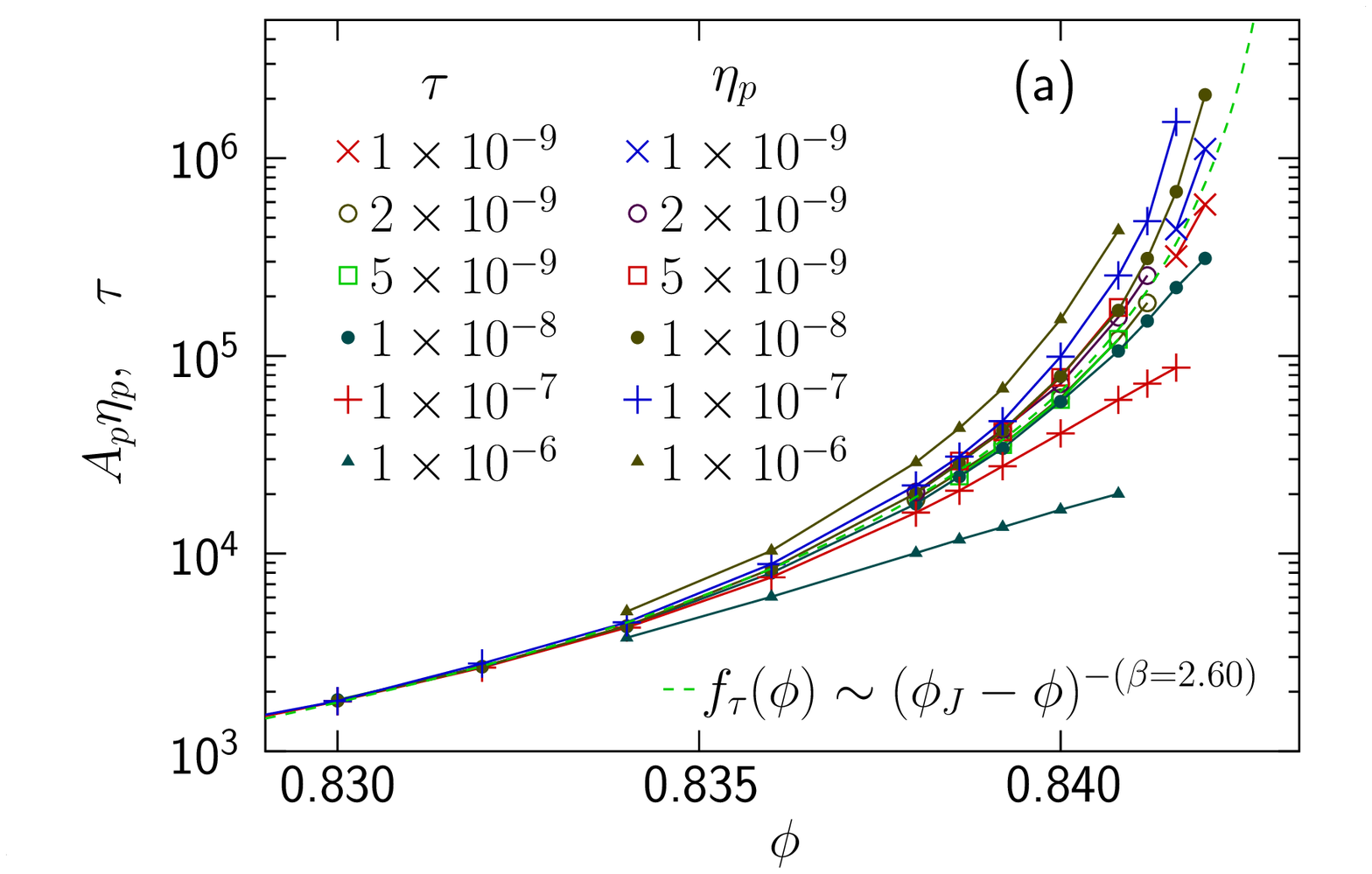}
  \includegraphics[bb=11 324 532 654, width=7cm]{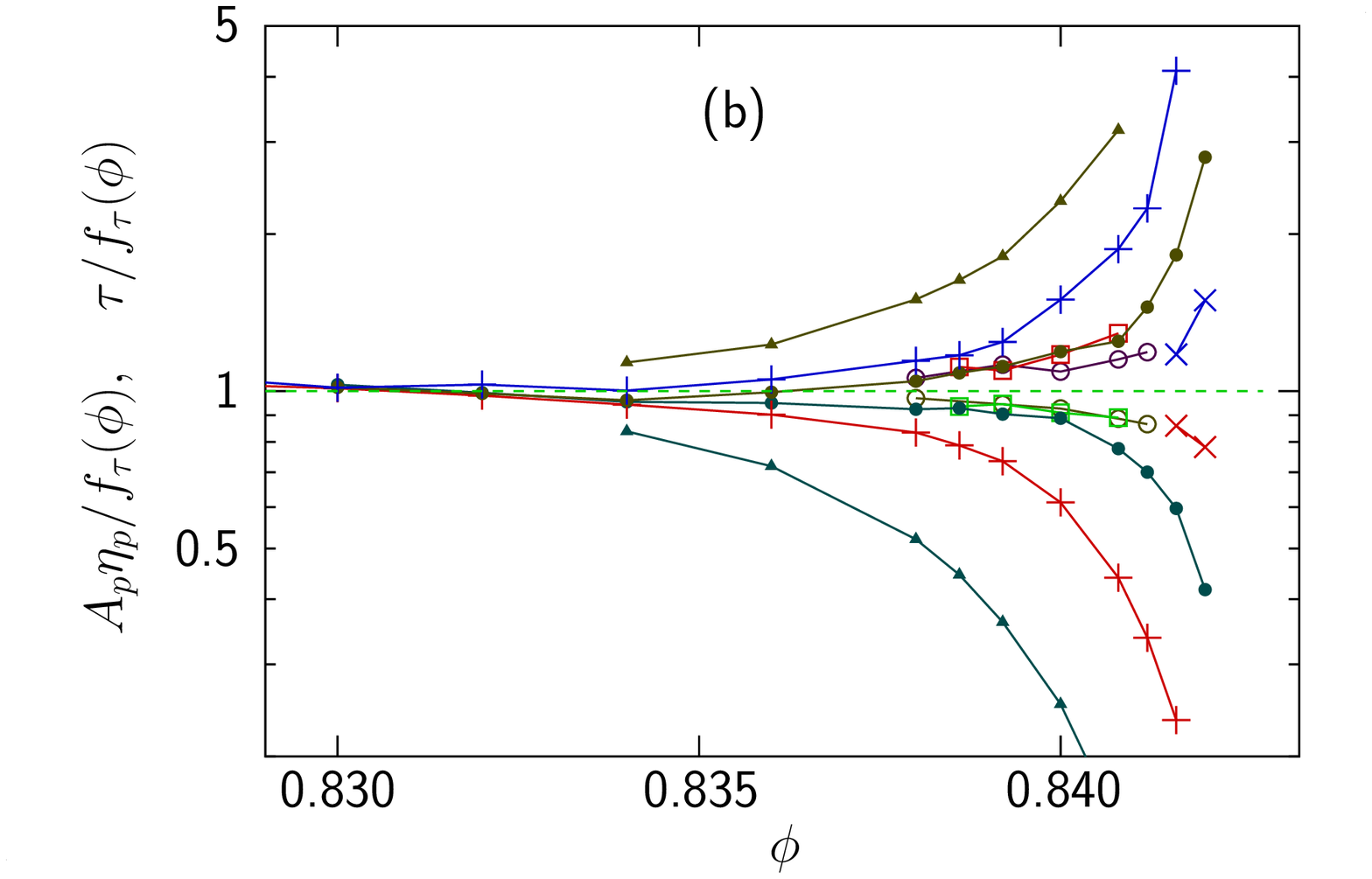}
  \caption{Comparison of $\tau$ and $\eta_p$ which, up to a constant prefactor, are
    expected to behave the same as $\gdot\to0$. Panel (a) shows the raw data, $\tau$, and
    $A_p\eta_p$, with the constant $A_p=36$. The data for large shear rates (solid
    triangles) are clearly different, but the respective points approach one another as
    $\gdot\to0$. The dashed line is $f_\tau(\phi)\sim(\phi_J-\phi)^{-\beta}$ with
    $\beta=2.60$. Panel (b) show the same data, but now divided by $f_\tau(\phi)$. The
    figure clearly suggests that the data should agree in the $\gdot\to0$ limit.}
  \label{fig:etap,tau}
\end{figure}

\subsection{Contact number}
\label{sec:zf}

\subsubsection{Relaxation time and contact number}

A key result from the study of static packings is that jamming in frictionless systems
occurs when the coordination number is $z = z_\mathrm{iso} \equiv 2D$, which is the number
needed for mechanical stability.\cite{Alexander:1998} This is however exact only in the
absence of rattlers---particles that are not locked up at a fixed position as they have
less than three contacts. To eliminate rattlers we follow \Ref{Lerner-PNAS:2012} and
repeatedly remove all particles with less than three contacts. After removing the
rattlers, $z_1$ is obtained as the average number of contacts of the remaining particles.

\begin{figure}
  \includegraphics[width=7cm]{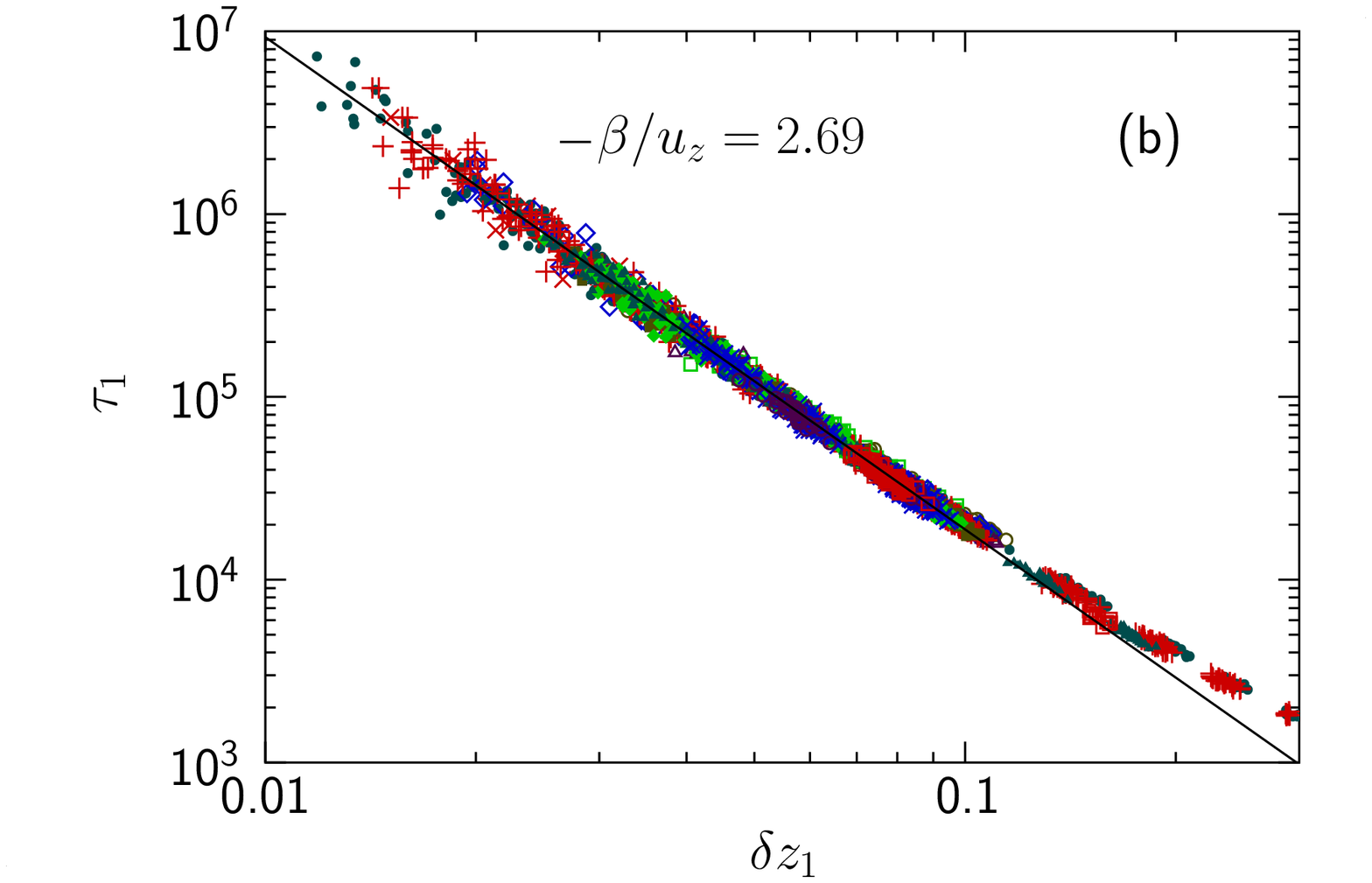}
  \includegraphics[width=7cm]{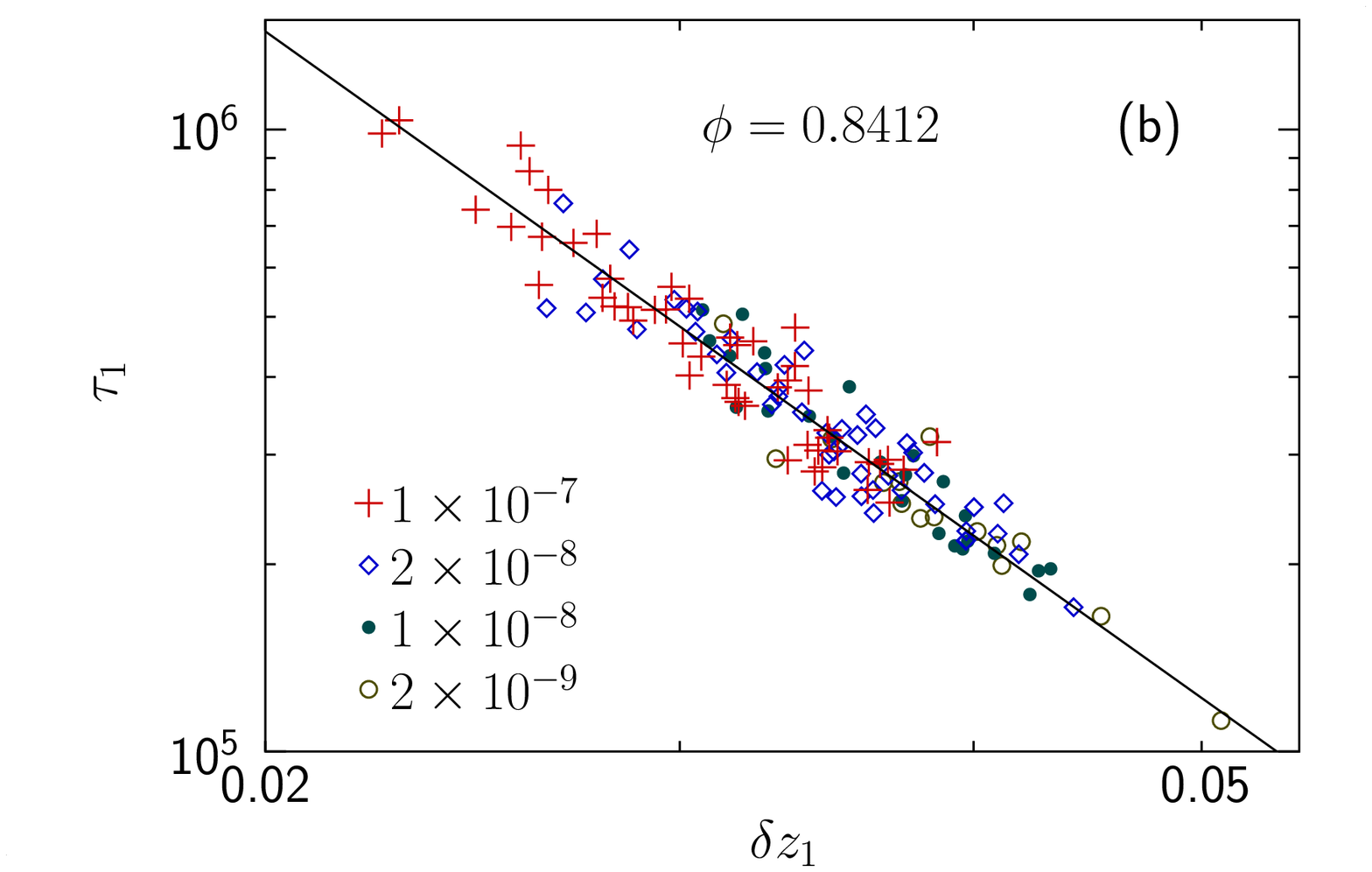}
  \caption{Corresponding values of $\tau_1$ and $\delta z_1$. Panel (a) shows 2719
    corresponding values of $\tau_1$ and $\delta z_1$. Each point is from a relaxation
    that gives both a relaxation time $\tau_1$ and a final configuration from which the
    contact number $z_1$ is determined. The relaxation time clearly depends algebraically
    on $\delta z_1$---the distance to isostaticity. A fit of all data with $\delta
    z_1<0.08$ (1625 points) gives the exponent $\beta/u_z=2.69$. Panel (b) is a zoom-in
    with a more restricted set of data: $\phi=0.8412$ and four different shear rates. This
    shows that the points for different initial shear rates fall on a single curve.}
  \label{fig:tau1-dz}
\end{figure}

Following Lerner \emph{et al.}\ \cite{Lerner-PNAS:2012} we show the individual
determinations, $\tau_1$ against $\delta z_1 \equiv z_\mathrm{iso} - z_1$ in
\Fig{tau1-dz}(a). The figure gives strong evidence for an algebraic relation. For the
vanishing of $\delta z_1$ we introduce $u_z$,
\begin{equation}
  \label{eq:dz-dphi}
  \delta z \sim (\phi_J-\phi)^{u_z}.
\end{equation}
Together with $\tau\sim(\phi_J-\phi)^{-\beta}$ this gives a relation between the
individual data points $\tau_1$ and $z_1$,
\begin{equation}
  \label{eq:tau1-dz1}
  \tau_1\sim(\delta z_1)^{-\beta/u_z},
\end{equation}
and a fit of our data gives the exponent $\beta/u_z=2.69$. Since there is a curvature in
the data that sets in around $\delta z_1=0.1$, only data with $\delta z_1<0.08$ were used
in the fit. This result appears to be especially robust since it is obtained from a very
simple fit of the raw data with no adjustable parameter. (Compare \Fig{tau-dphi} where a
determination of $\beta$ depends on the correct value of $\phi_J$.) Note also that there
is no need to restrict the data to small shear rates of the \emph{initial} simulation
stage. As shown in \Fig{tau1-dz}(b) data for different $\gdot$ do indeed fall on (or
spread around) the same line. The explanation for this seems to be that both $\tau_1$ and
$z_1$ are determined from configurations with almost vanishing overlaps, essentially in
the hard disk limit, independent of the initial shear rate. Together with
$\beta=2.70$\cite{Olsson_Teitel:gdot-scale} this suggests $u_z=1$ whereas the somewhat
smaller $\beta=2.58$ \cite{Olsson_Teitel:jam-HB} which would imply $u_z\approx 0.96$,
means that we cannot exclude the possibility that $u_z$ takes on a non-integral value.

% plo addon-tau1-dz1-8412.plo (also reads and executes tau-dz.plo)

Our result $\beta/u_z=2.69$ is in good agreement with \Ref{Lerner-PNAS:2012} who found
$\beta/u_z=1/0.38 = 2.63$. A more recent paper by the same authors\cite{DeGiuli:2014},
however, suggests $\beta/u_z=1/0.3\approx 3.3$ (their Fig.~5(c)). This new and lower
exponent ($0.3<0.38$) is due to a curvature in their data, bending over from a larger
slope for $\delta z>0.1$ to this lower slope for $\delta z<0.1$. This bending over at
$\delta z_1\approx 0.1$ is similar to our \Fig{tau1-dz}(a), though the slopes are
different. We cannot offer any explanation for this difference. (The effect in
\Fig{mean-tau-dz}(a) below, which also leads to a larger value of $\beta/u_z$, doesn't
seem to be applicable in that case.)

\begin{figure}
  \includegraphics[width=7cm]{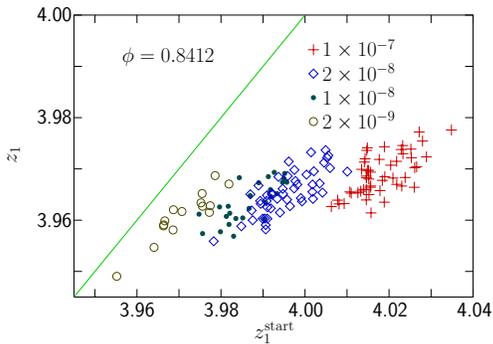}
  \caption{Change in contact number in the relaxation process. The figure shows contact
    numbers before and after the relaxation. The solid line is $z_1=z_1^\mathrm{start}$.
    The configurations are at density $\phi=0.8412$; the starting configurations are
    generated with four different initial shear rates. Both the initial
    $z_1^\mathrm{start}$ and $z_1$, obtained after the relaxation, are calculated after
    repeatedly removing all particles with less than three contacts.}
  \label{fig:z1-z1start}
\end{figure}

As mentioned above the contact numbers were determined from the relaxed configurations
with almost vanishing particle overlaps. To check if it would be possible to do a similar
analysis of the configurations before the relaxations, we have also determined the
corresponding starting values, $z_1^\mathrm{start}$, and to see how the relaxation process
changes the contact number \Fig{z1-z1start} shows the final contact number, $z_1$ against
the corresponding starting values, $z_1^\mathrm{start}$. These data are obtained for
$\phi=0.8412$, closely below $\phi_J$, and four different shear rates. From the figure we
may draw a few different conclusions: (1) The contact number always decreases in the
relaxation process. (2) This change is bigger for larger initial shear rates. (3) The
final $z_1$ decreases slowly with decreasing initial shear rate. (4) The contact number of
the starting configurations is sometimes above isostaticity,
$z_1^\mathrm{start}>z_\mathrm{iso}$ whereas $z_1$ is always below. This last point makes
clear that the analyses above, where the approach to jamming is seen by $z_1\to
z_\mathrm{iso}$ can not be used with $z_1^\mathrm{start}$; it is only $z_1$ obtained from
the relaxed configurations that approaches $z_\mathrm{iso}$ as jamming is approached.

% plo z1-z1start.plo

\subsubsection{Analysis of the CD$_0$ model}

\begin{figure}
  \includegraphics[width=7cm]{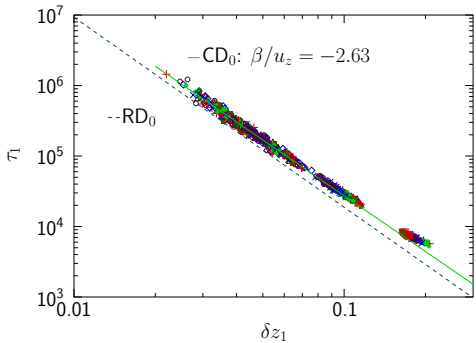}
  \caption{Determination of $\beta/u_z$ for the CD$_0$ model. By fitting data for $\delta
    z_1<0.08$ to \Eq{tau1-dz1} we determine $\beta/u_z=2.63$. We note that this is very
    close to $\beta/u_z=2.69$ of the RD$_0$ model.}
  \label{fig:CD-taup-dz}
\end{figure}

We have also applied the methods discussed above to the CD$_0$ model. These results are
from a rather limited number of relaxations and no data very close to jamming, but they
nevertheless give convincing results. \Fig{CD-taup-dz} shows $\tau_1$ vs $\delta z_1$ just
as in \Fig{tau1-dz}. The solid line, from fitting the data with $\delta z_1<0.08$, gives
the exponent $\beta/u_z=2.63$. We note that this is very close to $\beta/u_z=2.69$ of the
RD$_0$ model which gives support to the recent claim\cite{Vagberg_OT:jam-cdrd} that these
two models have the same critical behavior.  To facilitate a direct comparison, the
fitting line in \Fig{tau1-dz} is included as a dashed line in \Fig{CD-taup-dz}. The only
difference appears to be that the the relaxation time for the CD$_0$ model is about a
factor 1.5 larger than for the RD$_0$ model, for the same value of $\delta z_1$.

% cd /home/olsson/mcdata/hpc2n/jam-2014/CD-relax; plo tau1-dz1.plo

\subsubsection{Effect of large fluctuations}

\begin{figure}
  \includegraphics[width=7cm, bb=11 324 532 659]{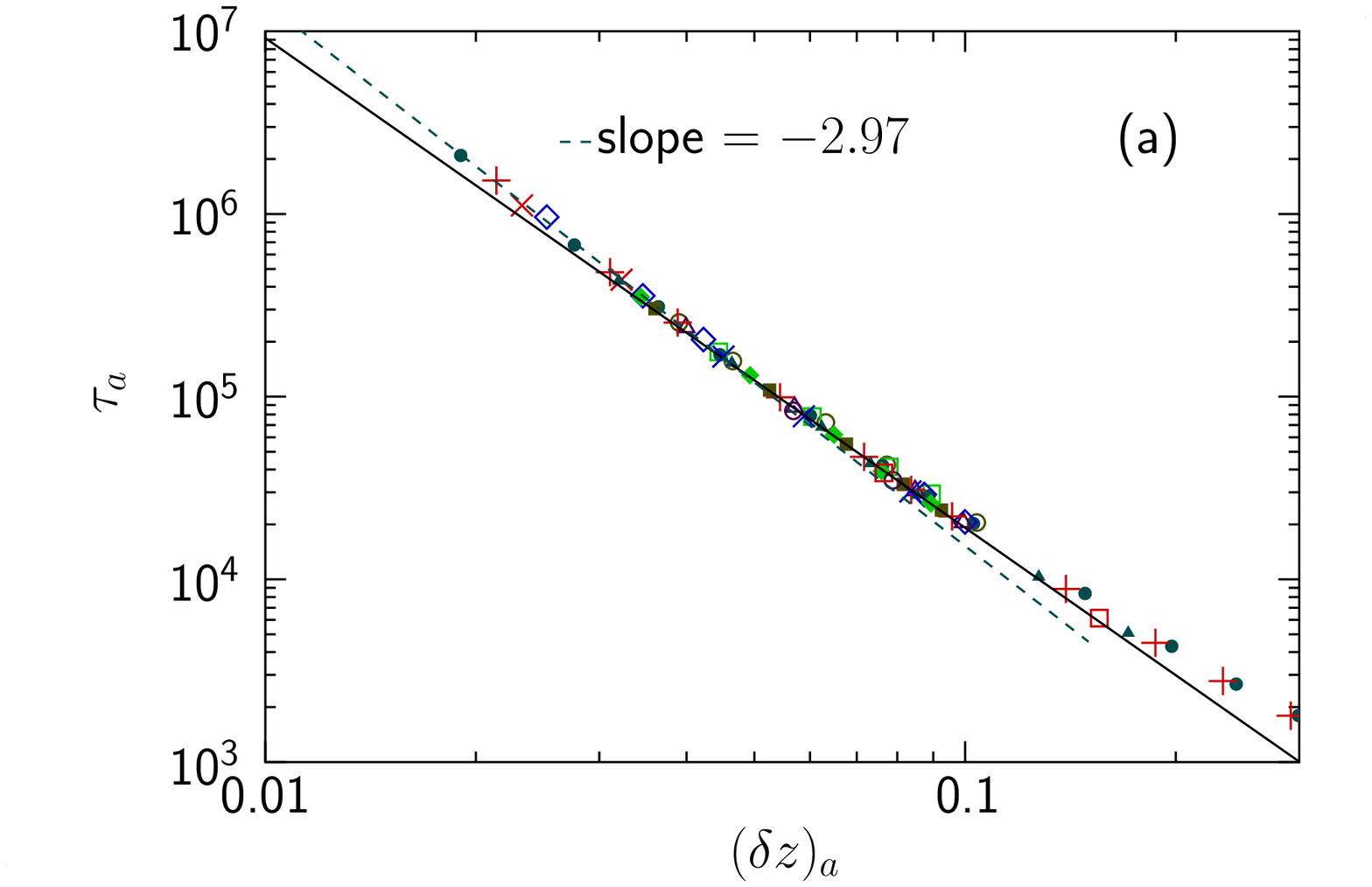}
  \includegraphics[width=7cm, bb=11 324 532 659]{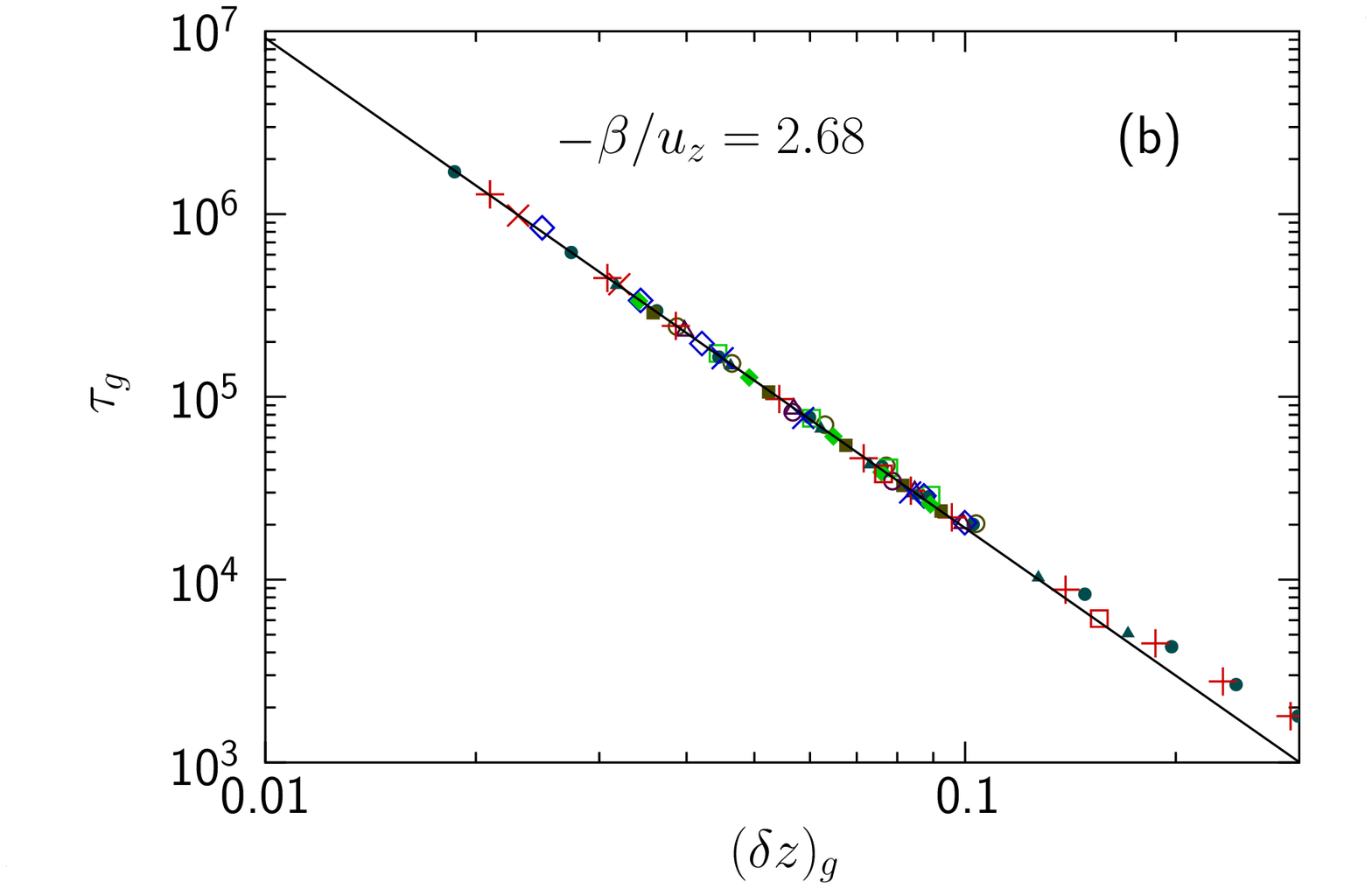}
  \caption{Mean values of $\tau_1$ and $\delta z_1$ determined in two different
    ways. Panel (a) shows the ordinary arithmetic mean values. For small $\delta z$ these
    points deviate clearly from the expected algebraic behavior. This phenomenon is due to
    the large spread of the data which appears close to jamming as is also described in
    conjuction with \Figs{a-g}. Panel (b) which shows the geometric means, $\tau_g$, and
    $(\delta z)_g$ of the points $(\tau_1, \delta z_1)$ in \Fig{tau1-dz}(a) for the same
    $\phi$ and $\gdot$. These points obey an algebraic behavior with the exponent
    $\beta/u_z=2.68$ in very good agreement with the analysis of the individual data
    points in \Fig{tau1-dz}(a).}
  \label{fig:mean-tau-dz}
\end{figure}

\Figure{tau1-dz} above displayed the individual data points $(\tau_1,\delta z_1)$, with
different symbols for different simulation parameters $\phi$, $\gdot$. An obvious way to
show the same thing in a less crowded figure, would be to determine the arithmetic means
of $\tau_1$ and $z_1$ for the different sets $(\phi,\gdot)$. We introduce the notation
$\tau_a$ and $(\delta z)_a$ for these arithmetic means. ($\tau_a$ is thus just the
ordinary average, $\tau$.)  This kind of data is shown in \Fig{mean-tau-dz}(a), and it
then turns out that the averaged data don't behave quite the same as the individual
points; the few points at the smallest $(\delta z)_a$ are now clearly off the solid
line. The reason for this is that the $\tau_1$ for a certain combination of $\phi$,
$\gdot$ are spread over a finite range of $\delta z$ and since there is a power law
relation between $\tau$ and $\delta z$, if one does the arithmetic average of this fixed
phi data, one gets a point that does not lie on the same curve.

However, it turns out that things work differently---all the
data fall on the line---when one instead plots the geometric means,
\begin{eqnarray}
  \label{eq:geom}
  \tau_g(\phi,\gdot) & = & \exp\left(\expt{\ln\tau_1^{(\phi,\gdot)}}\right),\\
  (\delta z)_g(\phi,\gdot) & = & \exp\left(\expt{\ln\delta z_1^{(\phi,\gdot)}}\right).
\end{eqnarray}
This data is shown in \Fig{mean-tau-dz}(b).

% again: plo tau-dz.plo

\begin{figure}
  \includegraphics[bb=50 328 354 654, width=4cm]{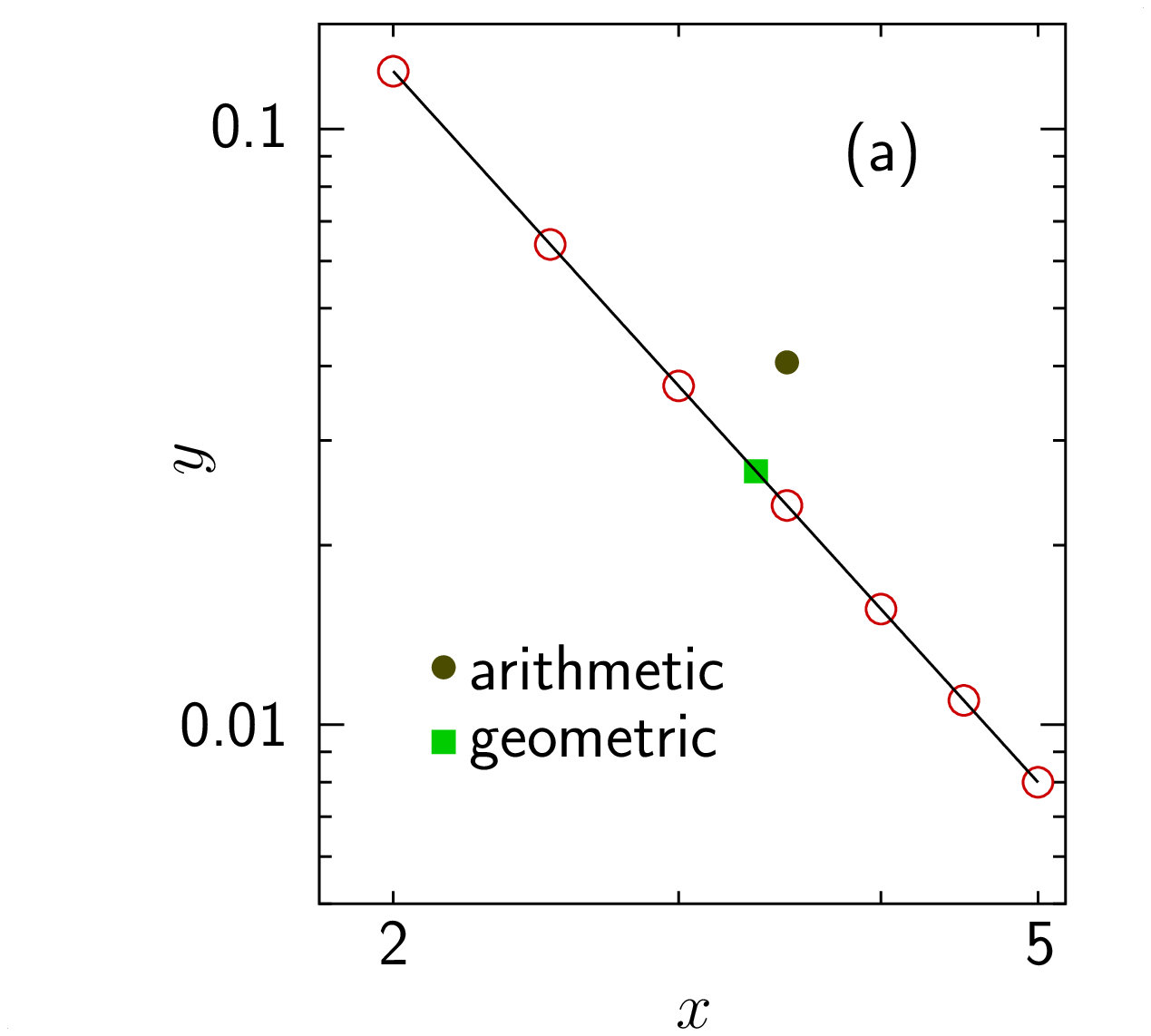}
  \includegraphics[bb=50 328 354 654, width=4cm]{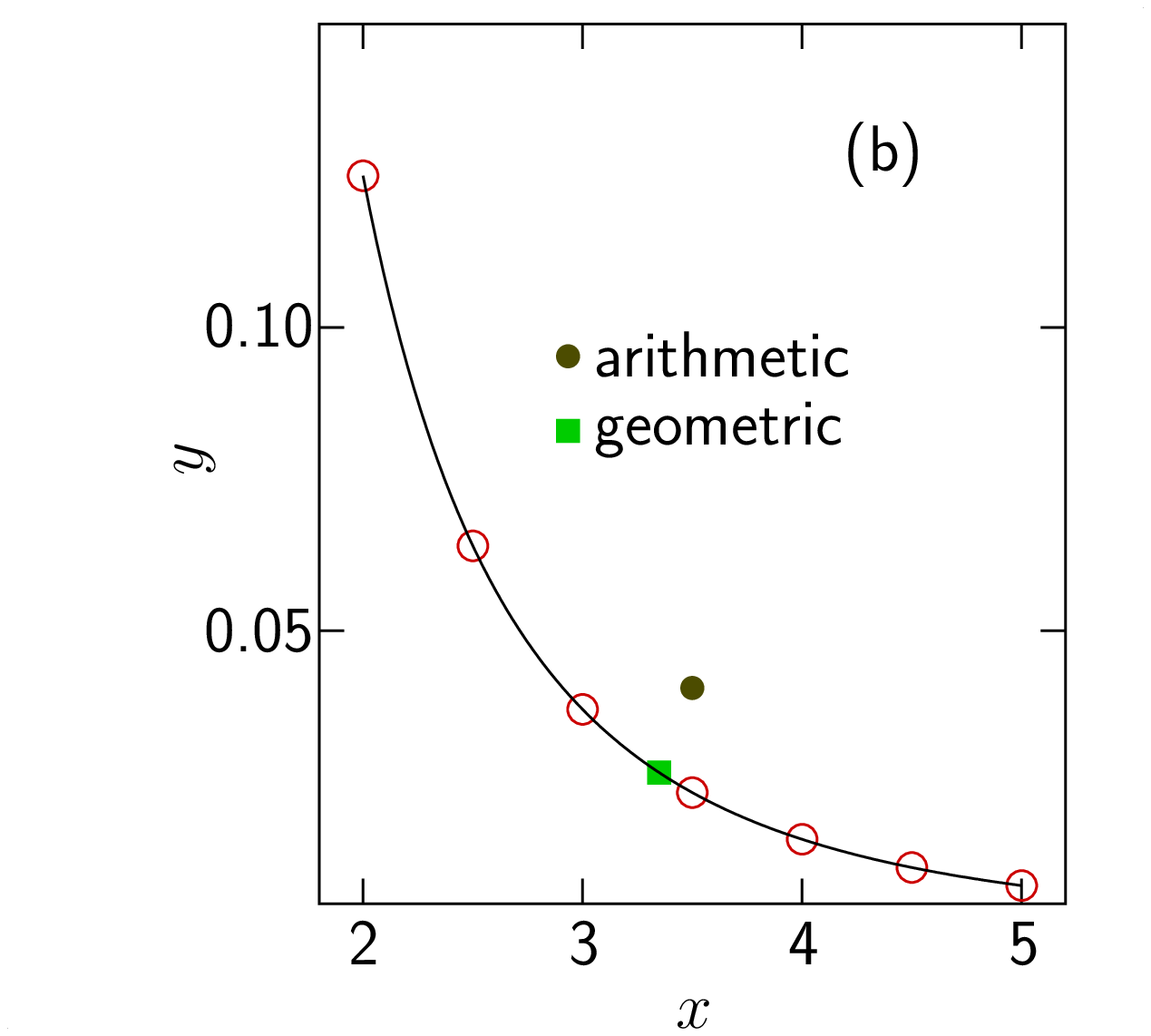}
  \caption{Illustration of the arithmetic mean and the geometric means for some points on
    the curve $y=x^{-3}$. From the figure with linear scale in panel (b) it is clear that
    one cannot expect the arithmetic mean to lie on top of the curve. As discussed in the
    text this effect only becomes important in cases where the relative variance is
    sizeable.}
  \label{fig:a-g}
\end{figure}

To illustrate what happens when one averages data with a power law relation, \Figs{a-g}
show the behavior of arithmetic and geometric means for some points on the line
$y=x^{-3}$, on logarithmic and linear scales, respectively. The points labelled
``arithmetic'' and ``geometric'' are the respective averages of the open circles in the
figures. In the left panel, which shows the data on logarithmic scales, the arithmetic
average is again, just as in \Fig{mean-tau-dz}(a), clearly off the line. Though this could
seem surprising, a plot with linear scales as in panel (b) directly shows that the
arithmetic average cannot lie on that line.

This effect is directly related to the big spread in the data around the average together
with a power different from one. With points $y_i=y_a(1+\delta_i)$ where $y_a$ is the
arithmetic mean and $\delta_i$ the relative deviation from this mean, the variance is
$\sigma_y^2 = \expt{y^2}-\expt{y}^2 = y_a^2 \expt{\delta_i^2}$. To second order in the
deviations, the geometric mean becomes
\begin{eqnarray*}
  \label{eq:hej}
  y_g & = & 
  \exp\left(\expt{\ln [y_a(1+\delta_i)]}\right) \\
  & \approx & y_a \exp\left(\expt{\delta_i-\delta_i^2/2}\right) \approx 
  y_a(1-\expt{\delta_i^2/2}),
\end{eqnarray*}
and the ratio of the two different averages becomes
\begin{equation}
  \label{eq:xg}
  \frac{y_g}{y_a} = 1-\frac{\sigma_y^2}{2y_a^2},
\end{equation}
which means that the effects discussed here are important only when the fluctuations in
the data are truly large.

\subsubsection{Finite size dependence}

% plo tau-dz-N-8380.plo (cd Finitesize; plo tau-dz-N-8380.plo)
\begin{figure}
  \includegraphics[width=7cm]{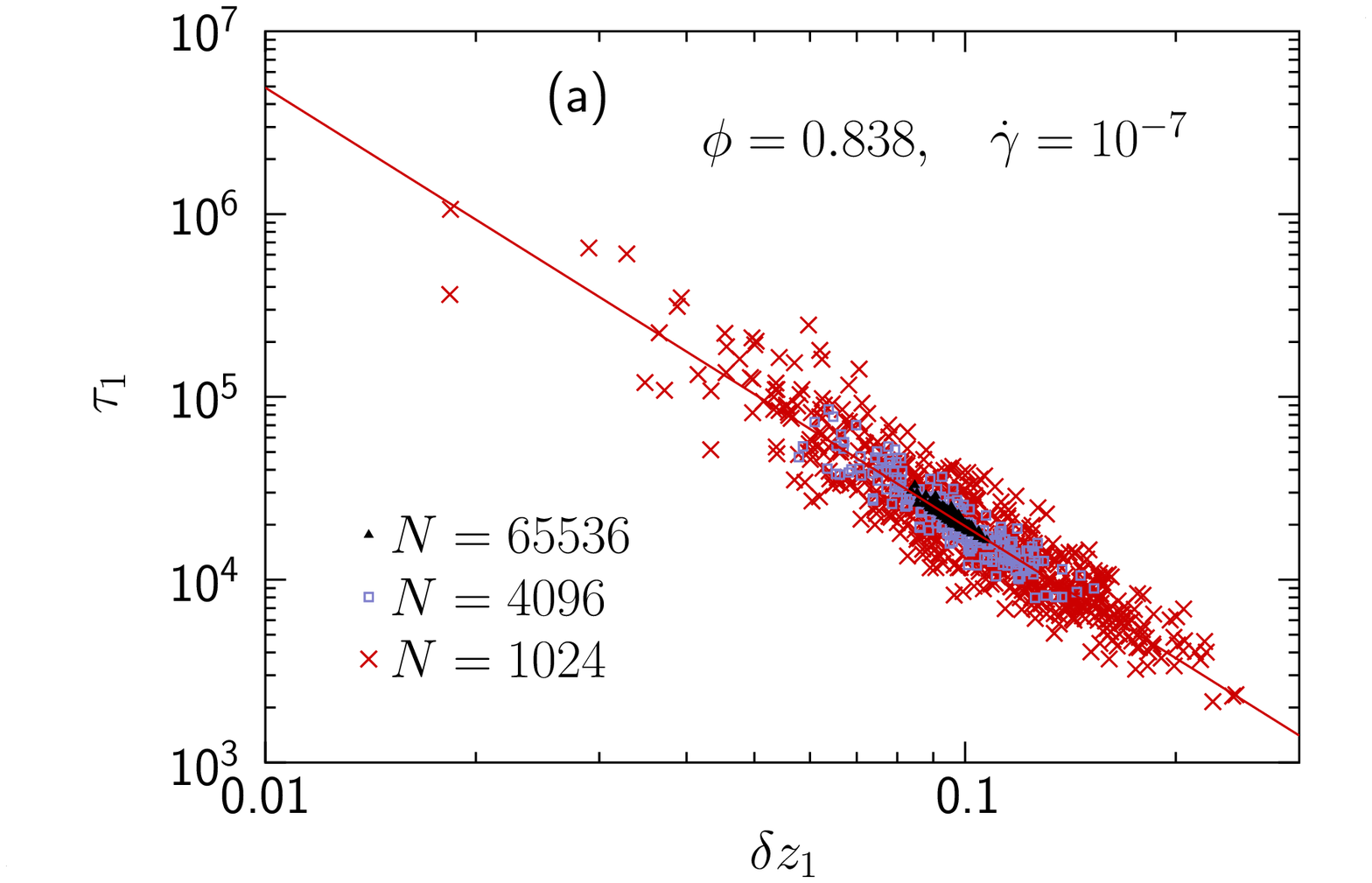}
  \includegraphics[width=7cm, bb=11 324 532 659]{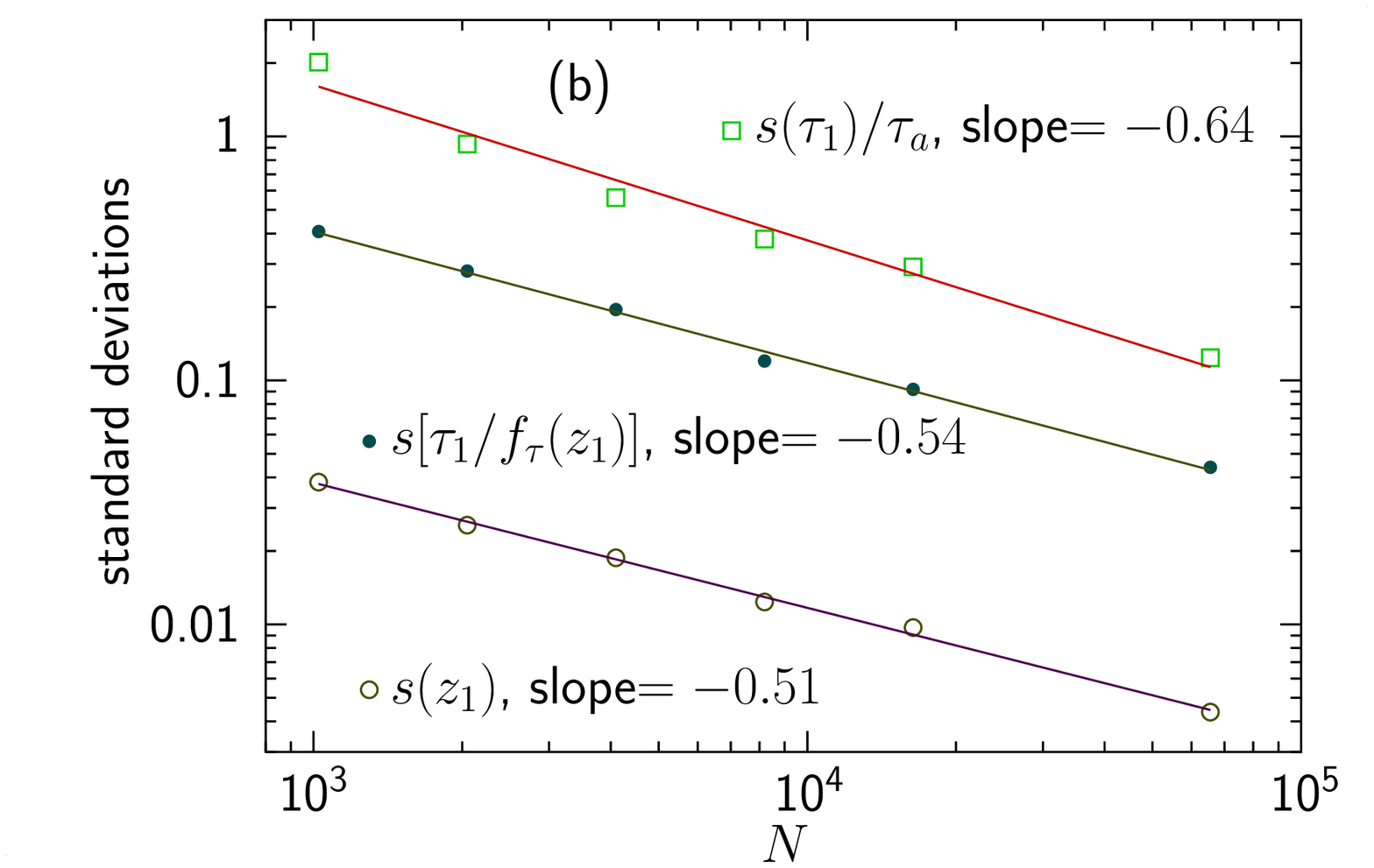}
  \caption{Finite size and the spread of the points $(\tau_1,\delta z_1)$ for $\phi=0.838$
    and initial shear rate $\gdot=10^{-7}$. Panel (a) which is $(\tau_1,\delta z_1)$ for
    three different system sizes shows that the points spread considerably more for smaller
    $N$. Panel (b) shows different quantitative measures of the spread of these data. The
    open circles are $s(z_1)$---the standard deviation of $z_1$. Open squares are
    $s(\tau_1)/\tau_a$. (The normalization by $\tau_a$ is to get quantities of the same
    order of magnitude). Solid dots are the standard deviation of $\tau_1/f_\tau(z_1)$
    which is the relative deviation of $\tau_1$ from the solid line in panel (a). Note
    that both the spread of $z_1$ and the spread around the solid line vanish as $1/\sqrt
    N$, as if the data were averages of $N$ independent samples.}
  \label{fig:tau1-dz1-N}
\end{figure}

We now examine the spread of $z_1$ and $\tau_1$, as in \Fig{tau1-dz}(b), around the solid
line, with special focus on how this spread depends on the finite system size. For the
finite size study we turn to a lower packing fraction, $\phi=0.838$. The reason for this
is that, closer to $\phi_J$ (e.g.\ at $\phi=0.840$) some configurations for smaller sizes
fail to reach zero energy in the relaxation step and get jammed with $z>z_\mathrm{iso}$,
and such events badly complicate the analysis. 

\Fig{tau1-dz1-N} which is $\tau_1$ vs $z_1$ for $\phi=0.838$, the initial shear rate
$\gdot=10^{-7}$, and the three sizes, $N=1024$, $N=4096$, and 65536, clearly shows that
these data spread more for smaller $N$.  Note that the data in \Fig{tau1-dz1-N} for all
different sizes have a common behavior, $\tau_1 \approx f_\tau(z_1) \equiv A_\tau (\delta
z_1)^{-b}$. The exponent $b=2.40$ is an effective exponent which differs from
$\beta/u_z=2.69$ (obtained in \Fig{tau1-dz}) since we here make use of data with larger
$\delta z_1$.

We introduce three different measures to characterize the spread of this data. Two
straightforward measures are $s(z_1)$ and $s(\tau_1)$ which are the standard deviations of
the data. Another measure is the spread of $\tau_1$ away from the line, i.e.\ the value
predicted from the known $z_1$, $s[\tau_1/f_\tau(z_1)]$. These three quantities are shown
in \Fig{tau1-dz1-N}(b) for number of particles ranging from $N=1024$ through 65536. To
interpret this data we first recall that the standard deviation of averages of $N$
independent samples is $\sim N^{-1/2}$. We find that both $s(z_1)$ and $s(\tau_1)/\tau_a$
vanish with the exponents $-0.54$ and $-0.51$ in excellent agreement with this
expectation. For $s[\tau_1/f_\tau(z_1)]$ we find a somewhat more complicated behaviour
with a larger exponent, $-0.64$, and a questionable fit to the data. Taken together our
data suggest an interpretation where both the spread of $z_1$ and the spread of $\tau_1$
around $f_\tau(z_1)$ are controlled by independent simple stochastic processes.

\section{Discussion}

The relaxation dynamics around the jamming transition has been studied before, but with a
rather different approach \cite{Hatano:2009}: the configurations were first generated
randomly, then relaxed to a zero-energy state with the conjugate gradient method, and
after that perturbed by a pure affine shear deformation. The relaxation time was then
determined from the relaxation of such initial states by fitting the shear stress to
$\sigma(\phi,t) \sim t^{-\alpha} e^{-t/\tau}$ with $\alpha=0.55(5)$, and the relaxation
time was found to diverge as $\tau\sim(\phi_J-\phi)^{-\zeta}$ with $\zeta=3.3(1)$. This
exponent is clearly bigger than our $\beta\approx2.7$. One possible explanation for this
difference is that we in the present study get data in the limit of vanishing shear rate
in the preparation step (i.e.\ $\gdot\to0$ in the steady state shearing), whereas they in
their work apply the pure shear deformation suddenly, which is more like a rapid
shearing. Indeed, as shown in \Fig{tau-dphi}(a) any given fixed shear rate would give too
large values for $\tau$ as one gets close to $\phi_J$, and from analyses of such data one
would expect to get too high values of the exponent for the divergence.

We finally want to stress two consequences of the presented results: We first stress that
the above results taken together suggest that $\tau$ is a fundamental quantity that
controls the overlap $\delta/\gdot$ and thereby is behind the divergence of other
quantites like $\eta_p$ and $\eta$. For a detailed argument we consider the $\gdot\to0$
limit where $\taudiss\approx\tau$ and the $N\to\infty$ limit where the spread of $z_1$ and
$\tau_1$ vanish. A given $\phi$ then leads to a well-defined $\delta z$ which in turn
implies a well-defined $\tau$ and $\taudiss\approx\tau$. With the additional assumption of
a given value for the dimensionless friction, $\mu\equiv\sigma/p$, power balance between
the input power $P_\mathrm{in} = \sigma\gdot\sim \mu\delta\gdot$ and the dissipated power
$P_\mathrm{diss} = E/\taudiss \sim \delta^2/\taudiss$ gives $\delta/\gdot
\sim\taudiss\mu$. This therefore provides a very direct link between the relaxation times
and $\eta_p\sim\delta/\gdot$.

Seconly, we note that the relaxation time $\tau$ we have defined here has a different
scaling exponent than does the time scale associated with rescaling the shear strain rate
$\gdot$.  From \Eq{p-divergence} for the $\gdot\to0$ limit and dimensional arguments one
would expect the deviations due to a finite $\gdot$ to scale as
\begin{equation}
  \label{eq:naive}
  \eta_p(\phi,\gdot)/ |\delta\phi|^{-\beta} \sim g(\gdot\tau) \sim
  g(\gdot/|\delta\phi|^\beta),\quad\mbox{(naive)},
\end{equation}
where the scaling function $\lim_{x\to0} g(x) =$ const (for the hard disk limit) and the
deviations being controlled by $\gdot\tau$.  This is however not the case. As shown in
\Eq{Ogdot-scale} the data scale with $g(\gdot/|\delta\phi|^{z\nu})$ where $z\nu=\beta+y$,
$y\approx 1.1$, which thus is clearly different from the behavior expected from
dimensional analysis. We hope to be able to return to this question elsewhere.

\section{Summary}

To summarize, we have done extensive two-step simulations, first shearing the system at
different constant shear rates and then stopping the shearing and letting the system
relax. At late times of this relaxation, both energy and pressure decay exponentially, and
we define the relaxation time, $\tau$, to be the time constant of the exponential decay of
the pressure. We similarly define the ``dissipation time'' from the initial decay
immediately after the shearing is turned off.

We then show that these two times behave very similarly when considering the limit of low
shear rates, but also that their respective shear rate dependencies are opposite. From the
expression for $\taudiss$, \Eq{taudiss}, it follows immediately that $\taudiss$ diverges with the
exponent $\beta$---the same divergence as for $\eta_p=p/\gdot$---and this is also
corroborated by the $\phi$-dependence of $\tau$ and $\taudiss$ in the small-$\gdot$ limit.

We also show that the relaxation time is directly related to the lowest vibrational
frequency of hard disk systems\cite{Lerner-PNAS:2012}, and, furthermore, that this
suggests a relation between $\tau$ and $\eta_p$, which should be valid in the small
$\gdot$ limit.  \Fig{etap,tau}, provide ample evidence that this actually is the case.

We then turn to a thorough study of the relation between the contact number and the
relaxation time.  The contact number is a key quantity in the field of jamming and we
follow \Ref{Lerner-PNAS:2012} and determine the contact number after removing
rattlers. With $\tau_1$ and $z_1$ from individual measurements, $\tau_1$ depends
algebraically on the distance from isostaticity $\delta z_1 = z_\mathrm{iso}-z_1$,
$\tau_1\sim (\delta z_1)^{\beta/u_z}$, with $\beta/u_z\approx 2.69$.

The same analysis applied to the CD$_0$ model gives essentially the same exponent,
$\beta/u_z\approx 2.63$, which provides additional evidence\cite{Vagberg_OT:jam-cdrd} that
the CD$_0$ and the RD$_0$ models are in the same universality class. We consider these
analysis to be especially robust as they are entirely straightforward and do not require
data obtained at very low shear rates.

We then turn to effects of the spread of the individual $\tau_1$ for a fixed set of
parameters $\phi$, $\gdot$, around its average. We first point out that the ordinary
arithmetic mean may be problematic and that a geometric mean actually in some respects
works better. We then consider the finite size effect where we find that the spread of
both the relaxation time and the coordination number go as $1/\sqrt N$, just as expected
for the statistics of $N$ independent variables.

I thank S. Teitel for many discussions and a critical reading of the manuscript.  This
work was supported by the Swedish Research Council Grant No.\ 2010-3725. Simulations were
performed on resources provided by the Swedish National Infrastructure for Computing
(SNIC) at PDC and HPC2N.  

\bibliography{j}
\bibliographystyle{apsrev4-1}
\end{document}